\documentclass[conference]{IEEEtran}
\IEEEoverridecommandlockouts
\usepackage{cite}
\usepackage{amsmath,amssymb,amsfonts}
\usepackage{algorithmic}
\usepackage{graphicx}
\usepackage{textcomp}
\usepackage{xcolor}

\newcommand{\vect}[1]{\mathbf{#1}}

\def\Htran{\mbox{\tiny $\mathrm{H}$}}
\def\Ttran{\mbox{\tiny $\mathrm{T}$}}
\def\CN{\mathcal{N}_{\mathbb{C}}} 
\def\imagunit{\mathsf{j}}

\def\BibTeX{{\rm B\kern-.05em{\sc i\kern-.025em b}\kern-.08em
    T\kern-.1667em\lower.7ex\hbox{E}\kern-.125emX}}
\begin{document}

\title{Performance Analysis for ISAC Systems with 1-bit DACs
\thanks{The work by M. B. Salman and E. Björnson was supported by the Knut and Alice Wallenberg Foundation. The work by \"O. T. Demir was supported by 2232-B International Fellowship for Early Stage Researchers Programme funded by the Scientific and Technological Research Council of T\"urkiye.}
}

\author{\IEEEauthorblockN{Murat Babek Salman$^{\dagger}$, Özlem Tuğfe Demir*, Emil Björnson$^{\dagger}$}
\IEEEauthorblockA{\textit{$^{\dagger}$School of Electrical Engineering and Computer Science, KTH Royal Institute of Technology, Kista, Sweden}}
\IEEEauthorblockA{\textit{*Department of Electrical-Electronics Engineering, TOBB University of Economics and Technology, Ankara, Türkiye} \\
mbsalman@kth.se, ozlemtugfedemir@etu.edu.tr, emilbjo@kth.se}

}

\maketitle

\begin{abstract}
Low-resolution quantization constrains the maximum achievable gains of multiple-input multiple-output (MIMO) systems. While the adverse effects and mitigation strategies have been thoroughly analyzed for communication systems, the impact of low-resolution quantization on integrated sensing and communication (ISAC) systems remains insufficiently explored in the existing literature. In this paper, we propose an analysis and design framework to investigate and mitigate the effects of 1-bit digital to analog converters (DACs) for ISAC systems. Firstly, an analytical sensing signal-to-noise ratio (SNR) expression is derived by using the Bussgang decomposition. Furthermore, two different methodologies are proposed to design a transmit waveform that satisfies both communication and sensing requirements simultaneously. The first method uses a separate constant modulus (CM) sensing signal since CM signals are known to be more robust to nonlinear distortion than orthogonal frequency division multiplexing (OFDM) modulated signals. The second method employs the squared-infinity norm Douglas-Rachford splitting (SQUID) approach to construct the transmit waveform using nonlinear quantized precoding. Finally, the performance of the proposed methods are validated via numerical simulations to indicate the complexity-performance tradeoff between two different methods.
\end{abstract}


\section{Introduction}
Within the landscape of sixth-generation (6G) communication technologies, integrated sensing and communication (ISAC) emerges as a cornerstone, addressing the escalating demands for heightened environmental awareness and novel use cases \cite{10217169}. An essential advantage of ISAC systems lies in their capacity to discern the location of potential eavesdroppers, thereby enhancing the security of communication channels \cite{10373185}. Moreover, ISAC propels the localization capabilities in Internet of Things (IoT) scenarios \cite{9737357}. These systems require combining communication and radar functionalities seamlessly, capitalizing on shared hardware infrastructure, spectrum, and temporal resources to deliver optimal performance, which makes the ISAC framework a potential solution \cite{10214237}. This unification of radar and communication tasks improves the processing and hardware efficiency and spectrum utilization of 6G networks. A pivotal enabler of ISAC systems in the 6G era is the incorporation of multiple-input multiple-output (MIMO) systems. Leveraging MIMO technology enriches the ISAC capabilities by harnessing spatial diversity and resolution, enabling concurrent communication and sensing operations with heightened reliability and efficacy \cite{10124714}. 

As the number of antennas at the base station (BS) increases, so does the complexity and cost of the hardware involved \cite{6891254}. One viable path forward is to enable large-scale MIMO by switching from traditional high-end components to low-cost components, to keep the total cost manageable.
In the downlink, this calls for using low-resolution digital-to-analog converters (DACs) \cite{7967843}. By employing such DACs, power consumption and costs can be significantly reduced at the price of the distorted received signals. However, nonlinear signal processing schemes have been developed to mitigate the effects of the low-resolution quantizers, which can support the required functionality and performance of the MIMO systems \cite{7967843,9780031,10210086}. 

In the literature, there are studies that consider the effects of low-resolution DACs and propose methods to mitigate their adverse impact. In \cite{9780031}, an analysis is performed to investigate system performance for MIMO radar with 1-bit DACs and analog-to-digital converters (ADCs). The detection and false alarm probability expressions are derived, and a waveform design method is proposed. Moreover, in \cite{10210086}, the effects of 1-bit digital beamforming are demonstrated, and it can be seen that 1-bit quantization both deteriorates the beampattern and spectrum of the transmitted waveform. 
The authors of \cite{10064130} explore the problem of target detection in large-scale MIMO systems employing low-resolution ADCs. However, these studies only focus on radar operation, but communication tasks are not considered. On the other hand, the effects of low-resolution quantization on the communication systems performance are thoroughly investigated, and nonlinear precoding methods are proposed to overcome the distortion due to quantization \cite{7967843,7894211,9096379,10043013}.
However, despite the growing interest in ISAC systems, there are a few works that consider low-resolution quantization in this context. For instance, a nonlinear precoding strategy to mitigate low-resolution quantization effects in ISAC systems is proposed in \cite{9399801}. This strategy is based on the same objective function outlined in \cite{7967843}, with additional constraints aimed at improving estimation accuracy. Furthermore, the optimization approach described in \cite{9399801} imposes constraints that prevent the use of efficient algorithms like the squared-infinity norm Douglas-Rachford splitting (SQUID) algorithm to solve the optimization problem. Another relevant contribution discussed in \cite{9724233} involves formulating an optimization problem to synthesize transmitted waveforms resembling linear radar waveforms. However, it is important to note that this algorithm's attempt to optimize the entire transmit sequence leads to significant computational complexity.

In this paper, we develop an analytical framework to assess the sensing signal-to-noise ratio (SNR) performance of an ISAC system based on the Bussgang decomposition. In addition, we introduce a waveform design methodology by using a dedicated constant modulus (CM) signal. In the proposed waveform design, the transmission power for the communication signal is calculated by assuming infinite-resolution DACs and the remaining power is assigned for the CM sensing signal. It is shown via numerical evaluations that by using a dedicated CM sensing signal, one can improve the sensing SNR and bit-error-rate (BER) performance. The obtained analytical SNR expression is verified via simulations. Lastly, an optimization framework is introduced to optimize sensing and symbol detection performance jointly. This is achieved by incorporating a beampattern penalty term alongside the received data symbol accuracy term, which enables the computationally efficient SQUID algorithm to be implemented. Lastly, performance improvement in both sensing and communication tasks, thanks to the proposed method, is shown via numerical evaluations.

\section{System Model}
This paper considers the monostatic ISAC operation of a BS that simultaneously serves $K$ single-antenna data-receiving users while detecting a potential target at the angle $\varphi_0$. The BS is equipped with transmit and receive antenna arrays with $M_{\rm t}$ and $M_{\rm r}$ antennas, respectively, which are constructed as uniform linear arrays (ULAs) having half-wavelength spacing. The BS is equipped with 1-bit DACs, while all other hardware components are considered ideal to isolate and focus on the deteriorating effects of the 1-bit DACs. To perform the dual functions, both communication and sensing signals are transmitted from the BS simultaneously. The transmitted signal vector at the discrete time $n$ is denoted as ${\bf z}[n] = \mathcal{Q}({\bf x}[n])$, where ${\bf x}[n] \in \mathbb{C}^{M_{\rm t}}$ is the transmit waveform and $\mathcal{Q}(\cdot)$ is the quantization function: 
\begin{equation}
z_m[n] =  \sqrt{\frac{P}{2M_{\rm t}}} \operatorname{sgn}\left(\operatorname{Re}( x_m[n]  ) \right) + \imagunit  \sqrt{\frac{P}{2M_{\rm t}}} \operatorname{sgn}\left(\operatorname{Im}( x_m[n]) \right),
\end{equation}
where $\operatorname{sgn}(\cdot)$ is the sign function and  $P$ is the transmit power.

\subsection{Downlink transmission}
The received signal at user $k$ at time $n$  can be expressed as
\begin{equation}
    y_k[n] ={\bf h}_k^{\Htran} {\bf z}[n] + w_k[n],
\end{equation}
where ${\bf h}_k \in \mathbb{C}^{M_{\rm t} }$ is the communication channel for the $k^{\rm th}$ user, and $w_k[n]$ is the additive white Gaussian noise (AWGN) with zero mean and variance $\sigma_w^2$. In this study, we consider line-of-sight (LoS) user channels
\begin{equation}
    {\bf h}_k = \kappa_k {\bf a}_{M_{\rm t}}(\varphi_k),
\end{equation}
where $\kappa_k$ is the complex channel coefficient including the pathloss for user $k$, $\varphi_k$ is the angle-of-departure (AoD) of user $k$, and $ {\bf a}_{M_{\rm t}}(\varphi) =  \left[ 1,e^{-\imagunit\pi \sin(\varphi)},\cdots, e^{-\imagunit\pi (M_{\rm t}-1)\sin(\varphi)} \right]^{\Ttran}$ is the array steering vector. This study considers line-of-sight channels for user channels; however, the proposed analysis framework is valid for all channel types. We assume perfect channel state information (CSI) is available at the BS.
From the received signal, the intended symbol estimate ${\hat{s}}_k[n]$ can be obtained at the user as
\begin{equation}
    {\hat{s}}_k[n] = \beta_k y_k[n]
\end{equation}
where $\beta_k \in \mathbb{R}$ represents the scaling factor applied by user $k$ before data detection takes place. In this paper, we consider two different approaches to form the transmit waveform ${\bf x}[n]$. The first approach is to design the transmit waveform by assuming infinite-resolution DACs. Under this assumption, the transmit waveform can be represented as
\begin{equation}
    {\bf x}[n] = {\bf P} {\bf s}[n] + {\bf p}_0 s_0[n],
\end{equation}
where ${\bf P} \in \mathbb{C}^{M_{\rm t} \times K}$ is the precoding matrix for the communication waveform, ${\bf s}[n] = [s_1[n],\ldots,s_K[n]]^{\Ttran}$ is the intended signal vector for the communication users with $\mathbb{E}\{{\bf s}[n]{\bf s}^{\Htran}[n]\}={\bf I}_K$, and ${\bf p}_0$ is the beamforming vector for the sensing signal $s_0[n]$. The sensing signal is conventionally selected to be a constant envelope signal $|s_0[n]| = 1$. Zero-forcing (ZF) is chosen to obtain the precoding matrix as \cite{7967843}

\begin{equation}
    {\bf P} = \frac{1}{\lambda} {\bf H} \left( {\bf H}^{\Htran} {\bf H} \right)^{-1},
\end{equation}
where $ \lambda = \frac{1}{\sqrt{\rho_{\rm c}}}  \sqrt{ \operatorname{tr} ( ( {\bf H}^{\Htran} {\bf H} )^{-1} ) }$, ${\bf H} = [{\bf h}_1, \dots, {\bf h}_K]$, and $\rho_{\rm c}$ is the transmit power dedicated to communication signals. The beamforming vector for the sensing signal is designed to illuminate the search direction; however, it is projected to the nullspace of the user channels in order to avoid interference as
\begin{equation}
    {\bf p}_0 = \frac{\sqrt{\rho_{\rm s}} \left({\bf I}_{M_ {\rm t}} - {\bf H} \left( {\bf H}^{\Htran} {\bf H} \right)^{-1} {\bf H}^{\Htran} \right) {\bf a}_{M_{\rm t}}(\varphi_0)}{\sqrt{ {\bf a}_{M_{\rm t}}^{\Htran}(\varphi_0)  \left({\bf I}_{M_ {\rm t}} - {\bf H} \left( {\bf H}^{\Htran} {\bf H} \right)^{-1} {\bf H}^{\Htran} \right) {\bf a}_{M_{\rm t}}(\varphi_0)}}  ,
\end{equation}
where $\rho_{\rm s}$ is the dedicated sensing power, which can be computed as $\rho_{\rm s} = P-\rho_{\rm c}$. After quantization, we can represent the transmitted signal by using the \emph{generalized} Bussgang decomposition \cite{9307295} as

\begin{equation}
    {\bf z}[n] = {\bf B}{\bar {\bf{P}}} {\bar {\bf{s}}} [n]+ {\bf d}[n],
\end{equation}
where $\bf B $ is the Bussgang gain matrix, $\Bar{{\bf s}}[n] \triangleq [{\bf s}^{\Ttran}[n] s_0[n]]^{\Ttran}$, ${\bar {\bf{P}}} \triangleq [{\bf P}, {\bf p}_0]$, and ${\bf d}[n]$ is the distortion vector. By using the generalized Bussgang decomposition, the Bussgang matrix $\bf B$ can be obtained as ${\bf B} = {\bf C}_{zx} {\bf C}_{xx}^{-1}$, where ${\bf C}_{zx} = \mathbb{E} [{\bf z}[n] {\bf x}^{\Htran}[n]]$ and ${\bf C}_{xx} = \mathbb{E} [{\bf x}[n] {\bf x}^{\Htran}[n]]$, and the correlation matrix of the distortion vector can be obtained as ${\bf C}_{dd} =\mathbb{E} [{\bf d}[n] {\bf d}^{\Htran}[n]] ={\bf C}_{zz} - {\bf B} {\bf C}_{xx} {\bf B}^{\Htran}$. Assuming complex Gaussian input signals thanks to inverse fast Fourier transformation (IFFT) at the OFDM transmitter, the Bussgang gain matrix and ${\bf C}_{zz}$ can be expressed in closed forms as \cite{7967843} 
\begin{equation}
    {\bf B} = \sqrt{\frac{2P}{\pi M_{\rm t}}} \operatorname{diag}(\bar{{\bf P}} \bar{{\bf P}}^{\Htran})^{-1/2},
\end{equation}
\begin{equation}
    \begin{split}
        &{\bf C}_{zz} =\frac{P}{\pi M_{\rm t}} \sin^{-1}\left( \operatorname{diag}(\bar{\bf P} \bar{\bf P}^{\Htran})^{-1/2} \Re( \bar{\bf P} \bar{\bf P}^{\Htran}) \operatorname{diag}(\bar{\bf P} \bar{\bf P}^{\Htran})^{-1/2} \right)\\
        &+ \imagunit\frac{P}{\pi M_{\rm t}} \sin^{-1}\left( \operatorname{diag}(\bar{\bf P} \bar{\bf P}^{\Htran})^{-1/2} \Im( \bar{\bf P} \bar{\bf P}^{\Htran}) \operatorname{diag}(\bar{\bf P} \bar{\bf P}^{\Htran})^{-1/2} \right). \label{eq:Czz}
    \end{split}
\end{equation}
Then, we can write the received signal for all communication users as
\begin{equation}
\begin{split}
    {\bf y}_{\rm c}[n] =&  {\bf H}^{\Htran}{\bf B}{\bar {\bf{P}}} {\bar {\bf{s}}} [n] + {\bf H}^{\Htran} {\bf d}[n]  + {\bf w}_{ \rm c}[n],
\end{split}
\end{equation}
where ${\bf y}_{ \rm c}[n]=[y_1[n] ,  \cdots , y_K[n]]^{\Ttran}$ and ${\bf w}_{\rm c}[n]=[w_1[n] , \cdots , w_K[n]]^{\Ttran}$.
Note that one can define the communication SNR for user $k$ as ${\rm SNR}_k = \rho_{\rm c} |\kappa_k|^2/\sigma_w^2$.

\subsection{Sensing Receiver}
The echo signal received by the sensing array at the BS can be written as
\begin{align}
    {\bf y}_{\rm s}[n] = c{\bf a}_{M_{\rm r}}(\varphi_0) {\bf a}_{M_{\rm t}}^{\Htran}(\varphi_0){\bf z}[n-\tau]+{\bf w}_{\rm s}[n], 
\end{align}
for $n=0,\ldots,L-1$,
where $c\sim \CN(0,\sigma^2_{\rm RCS} |\kappa_0|^2)$ is the radar cross section (RCS), including the two-way pathloss, ${\bf a}_{M_{\rm r}}(\varphi_0)=\left[ 1,e^{-\imagunit\pi \sin(\varphi_0)},\cdots, e^{-\imagunit\pi (M_{\rm r}-1)\sin(\varphi_0)} \right]^{\Ttran}$ is the channel of the echo signal, 
${\bf w}_{\rm s}[n]\sim\mathcal{N}_{\mathbb{C}}({\bf 0},\sigma_w^2{\bf I}_{M_{\rm r}})$ is the independent noise vector added to echo signal at the BS, $L$ is the block length for OFDM symbol during the interval RCS is assumed to be constant as in Swerling I model, and $\tau$ is the discrete-time equivalent delay of the echo signal. To maximize the sensing SNR, a matched filter (MF) is applied to the received signal as
\begin{equation}
    {{\bf Y}}[\hat{\tau}]= \sum_{n=0}^{L-1} {\bf y}_{\rm s}[n] \Bar{{\bf s}}^{\Htran}[n-\hat{\tau}], \label{RadarRx}
\end{equation}
where ${{\bf Y}}[\hat{\tau}] \in \mathbb{C}^{ M_{\rm t} \times (K+1) }$ is the MF output at delay index $\hat{\tau}$, corresponding to the delay at which the target is being searched. Note that the quantized transmit signal is known by the receiver unit thanks to the monostatic setup. Then, we can express MF output as 
\begin{equation}
\begin{split}
    {{\bf Y}}[\hat{\tau}]& = c{\bf a}_{M_{\rm r}}(\varphi_0) {\bf a}_{M_{\rm t}}^{\Htran}(\varphi_0) \sum_{n=0}^{L-1} {\bf z}[n-\tau]\Bar{{\bf s}}^{\Htran}[n-\hat{\tau}] \nonumber\\
    &\quad + \sum_{n=0}^{L-1}{\bf w}_{\rm s}[n] \Bar{{\bf s}}^{\Htran}[n-\hat{\tau}]\\
    &\approx cL{\bf a}_{M_{\rm r}}(\varphi_0) {\bf a}_{M_{\rm t}}^{\Htran}(\varphi_0){\bf B}{\Bar{\bf P} } \nonumber\\
    &\quad+ c{\bf a}_{M_{\rm r}}(\varphi_0) {\bf a}_{M_{\rm t}}^{\Htran}(\varphi_0) {\bf D}_{\hat{\tau}}  + {\bf W}_{\hat{\tau}} ,
\end{split}
\end{equation}
where ${\bf D}_{\hat{\tau}} \triangleq \sum_{n=0}^{L-1} {\bf d}[n - \tau] \Bar{{\bf s}}^{\Htran}[n-\hat{\tau}]$, ${\bf W}_{\hat{\tau}} \triangleq \sum_{n=0}^{L-1} {\bf w}_{\rm s}[n] \Bar{{\bf s}}^{\Htran}[n-\hat{\tau}]$, and we used the common assumption of $\sum_{n=0}^{L-1} \Bar{{\bf s}}[n-\tau] \Bar{{\bf s}}^{\Htran}[n-\hat{\tau}] \approx L {\bf I}_{K+1} $. In order to simplify the expression we further consider $\tilde{{\bf a}} \triangleq {\bf a}_{M_{\rm t}}^{\Htran}(\varphi_0) {\bf B}{\Bar{\bf P} }$ as
\begin{equation}
\begin{split}
    \tilde{{\bf a}} &={\bf a}_{M_{\rm t}}^{\Htran}(\varphi_0) {\bf B} [{\bf p}_1, \, \cdots, {\bf p}_K, \, {\bf p}_0 ],\\
    & = [\sqrt{\rho_{\rm c}} g_1, \, \cdots ,\sqrt{\rho_{\rm c}} g_K, \, \sqrt{\rho_{\rm s}} g_0],
\end{split}    
\end{equation}
where $g_k = {\bf a}_{M_{\rm t}}^{\Htran}(\varphi_0) {\bf B} \Tilde{{\bf p}}_k$, $\Tilde{{\bf p}}_k \triangleq \frac{{\bf p}_k}{\sqrt{\rho_{\rm c}}} $, for $k=1,\ldots,K$ and $\Tilde{{\bf p}}_0 \triangleq \frac{{\bf p}_0}{\sqrt{\rho_{\rm s}}} $ . Then, the MF output can be vectorized and simplified as
\begin{align}
    \operatorname{vec}( {{\bf Y}}[\hat{\tau}]) &= cL ({\bf I}_{K+1} \otimes {\bf a}_{M_{\rm r}}(\varphi_0))\tilde{{\bf a}}^{\Ttran} \nonumber\\ \nonumber
    &\quad+ \operatorname{vec}(c{\bf a}_{M_{\rm r}}(\varphi_0) {\bf a}_{M_{\rm t}}^{\Htran}(\varphi_0) {\bf D}_{\hat{\tau}} ) + \operatorname{vec}({\bf W}_{\hat{\tau}}) \\
 &= c {\bf v} + \operatorname{vec}(c{\bf a}_{M_{\rm r}}(\varphi_0) {\bf a}^{\Htran}_{M_{\rm t}}(\varphi_0) {\bf D}_{\hat{\tau}} ) + \operatorname{vec}({\bf W}_{\hat{\tau}}),
\end{align}
where
${\bf v} = L\operatorname{bdiag}( \sqrt{\rho_{\rm c} }g_1{\bf a}_{M_{\rm r}}(\varphi_0), \cdots ,\sqrt{\rho_{\rm c} }g_K{\bf a}_{M_{\rm r}}(\varphi_0), \\  \sqrt{\rho_{\rm s} }g_0{\bf a}_{M_{\rm r}}(\varphi_0))$. In order to maximize the sensing SNR, MF is applied as
\begin{equation}
    \begin{split}
        \hat{y}[\hat{\tau}] &= {\bf v}^{\Htran} \operatorname{vec}( {{\bf Y}}[\hat{\tau}])\\
        & = c ||{\bf v}||^2 + {\bf v}^{\Htran}  \operatorname{vec}(c{\bf a}_{M_{\rm r}}(\varphi_0) {\bf a}_{M_{\rm t}}^{\Htran}(\varphi_0) {\bf D}_{\hat{\tau}} ) \\
        &\quad+ {\bf v}^{\Htran} \operatorname{vec}({\bf W}_{\hat{\tau}}). \label{MFoutput}
    \end{split}
\end{equation}
By using \eqref{MFoutput}, the sensing SNR (${\rm sSNR}$) can be written as
\begin{equation}
    \begin{split}
        \rm{sSNR} = \frac{|\kappa_0|^2\sigma_{\rm RCS}^2 \left(|| {\bf v}||^4 +   {\bf v}^{\Htran} \mathbb{E} \left[ \operatorname{vec}( \bar{{\bf D}}_{\hat{\tau}})  \operatorname{vec}( \bar{{\bf D}}_{\hat{\tau}})^{\Htran} \right] {\bf v}\right)}{{\bf v}^{\Htran} \operatorname{vec}({\bf W}_{\hat{\tau}})\operatorname{vec}({\bf W}_{\hat{\tau}})^{\Htran}{\bf v}},
    \end{split}
\end{equation}
where we have defined $\bar{{\bf D}}_{\hat{\tau}}={\bf a}_{M_{\rm r}}(\varphi_0) {\bf a}_{M_{\rm t}}^{\Htran}(\varphi_0) {\bf D}_{\hat{\tau}}$
and distortion is considered as a desired signal since its existence is beneficial for the subsequent hypothesis testing since it only appears if the target exists. Note that ${\rm sSNR}$ is a reasonable metric that indicates sensing reliability. Since the elements of the noise vector are independent and identically distributed (i.i.d.), the denominator can be simplified as ${\bf v}^{\Htran} \operatorname{vec}({\bf W}_{\hat{\tau}})\operatorname{vec}({\bf W}_{\hat{\tau}})^{\Htran}{\bf v} = L\sigma_w^2 || {\bf v}||^2$. The expression can be simplified further by considering the expectation ${\bf \Sigma} \triangleq \mathbb{E} \left[ \operatorname{vec}( \bar{{\bf D}}_{\hat{\tau}})  \operatorname{vec}( \bar{{\bf D}}_{\hat{\tau}})^{\Htran} \right]$ as
\begin{equation}
\begin{split}
    & {\boldsymbol{\Sigma}} =\sum_{n=0}^{L-1}\sum_{n'=0}^{L-1} \mathbb{E} \left[\begin{bmatrix} {\bf a}_{M_{\rm r}}(\varphi_0)  \tilde{ d}[n - \tau]  s_1^{*}[n-\hat{\tau}] \\ \vdots \\  {\bf a}_{M_{\rm r}}(\varphi_0)   \tilde{ d}[n - \tau]  s_K^{*}[n-\hat{\tau}] \\  {\bf a}_{M_{\rm r}}(\varphi_0)  \tilde{ d}[n - \tau]  s_0^{*}[n-\hat{\tau}] \end{bmatrix}\right.\\ \times &\left.\begin{bmatrix} {\bf a}_{M_{\rm r}}(\varphi_0)  \tilde{ d}[n' - \tau]  s_1^{*}[n'-\hat{\tau}] \\ \vdots \\  {\bf a}_{M_{\rm r}}(\varphi_0)  \tilde{ d}[n' - \tau]  s_K^{*}[n'-\hat{\tau}] \\  {\bf a}_{M_{\rm r}}(\varphi_0) \tilde{ d}[n' - \tau]  s_0^{*}[n'-\hat{\tau}] \end{bmatrix}^{\Htran} \right],
\end{split}
\end{equation}
where $\tilde{d}[n-\tau]= {\bf a}^{\Htran}_{M_{\rm t}}(\varphi_0) {\bf d} [n-\tau]$. Note that $ {\boldsymbol{\Sigma}}$ is a block diagonal matrix since different symbols are independent; hence, its $i^{th}$ block matrix can be written as
\begin{align} \nonumber
    {\boldsymbol{\Sigma}}_i &= {\bf a}_{M_{\rm r}}(\varphi_0) \sum_{n=0}^{L-1}  \mathbb{E} \left[   \left|\tilde{ d}[n - \tau]\right|^2 \left|\Bar{{ s}}_i[n-\hat{\tau}]\right|^2  \right] {\bf a}_{M_{\rm r}}^{\Htran}(\varphi_0)\\ \nonumber
    & = {\bf a}_{M_{\rm r}}(\varphi_0) {\bf a}_{M_{\rm r}}^{\Htran}(\varphi_0)\sum_{n=0}^{L-1}  \mathbb{E} \left[ \left|\tilde{ d}[n - \tau]\right|^2\right]  \mathbb{E} \left[ |\Bar{{ s}}_i[n-\hat{\tau}]|^2\right]\\
    &=L \sigma_{\Tilde{d}}^2{\bf a}_{M_{\rm r}}(\varphi_0){\bf a}_{M_{\rm r}}^{\Htran}(\varphi_0),\label{Sigma_i}
\end{align}
where $\sigma_{\Tilde{d}}^2 \triangleq {\bf a}_{M_{\rm t}}^{\Htran}(\varphi_0) {\bf C}_{dd} {\bf a}_{M_{\rm t}}(\varphi_0)$, and we assumed that distortion and the desired signal are independent. Then, signal power \emph{thanks to} the distortion component can be written as
\begin{equation}
    \begin{split}
        {\bf v}^{\Htran}{\bf \Sigma} {\bf v} = \sigma_{\tilde d}^2 L^3 \rho_{\rm c} M_{\rm r}^2 \sum_{k=1}^K  |g_k|^2  + \sigma_{\tilde d}^2L^3\rho_{\rm s} M_{\rm r}^2|g_0|^2 .
    \end{split}
\end{equation}
Lastly, we can obtain the norm square of the maximum ratio combiner $|| {\bf v}||^2$ as
\begin{equation}
    || {\bf v}||^2 = L^2 \rho_{\rm c} M_{\rm r}\sum_{k=1}^K |g_k|^2 + L^2 \rho_{\rm s} M_{\rm r} |g_0|^2.
\end{equation}
Finally, we can simplify the sensing SNR expression as in \eqref{SNRFinal}, on the top of the next page.
\begin{figure*}
    
\begin{equation}
\begin{split}
    {\rm sSNR} &= \frac{|\kappa_0|^2\sigma_{\rm RCS}^2 M_{\rm r}^2 L^4 \left(\rho_{\rm c} \sum_{k=1}^K |g_k|^2 + \rho_{\rm s} |g_0|^2 \right)^2 + |\kappa_0|^2\sigma_{\rm RCS}^2 \sigma_{\tilde d}^2 L^3 \rho_{\rm c} M_{\rm r}^2\sum_{k=1}^K  |g_k|^2  +|\kappa_0|^2\sigma_{\rm RCS}^2 \sigma_{\tilde d}^2L^3\rho_{\rm s}M_{\rm r}^2 |g_0|^2  }{\sigma_w^2L^3 \rho_{\rm c} M_{\rm r} \sum_{k=1}^K |g_k|^2 + \sigma_w^2L^3 \rho_{\rm s} M_{\rm r} |g_0|^2}\\
    &= \frac{|\kappa_0|^2\sigma_{\rm RCS}^2 M_{\rm r} L \left(\rho_{\rm c} \sum_{k=1}^K |g_k|^2 + \rho_{\rm s} |g_0|^2 \right)^2 +  |\kappa_0|^2\sigma_{\rm RCS}^2\sigma_{\tilde d}^2  \rho_{\rm c} M_{\rm r} \sum_{k=1}^K  |g_k|^2  +|\kappa_0|^2\sigma_{\rm RCS}^2 \sigma_{\tilde d}^2 \rho_{\rm s} M_{\rm r}|g_0|^2 }{\sigma_w^2 \rho_{\rm c}  \sum_{k=1}^K |g_k|^2 + \sigma_w^2\rho_{\rm s}  |g_0|^2}\\
    &= \frac{|\kappa_0|^2\sigma_{\rm RCS}^2 M_{\rm r} L \left(\rho_{\rm c} \sum_{k=1}^K |g_k|^2  + \rho_{\rm s} |g_0|^2 \right) + |\kappa_0|^2\sigma_{\rm RCS}^2 \sigma_{\tilde d}^2  M_{\rm r}}{\sigma_w^2}.\label{SNRFinal}
\end{split}
\end{equation}
\hrulefill
\end{figure*}

\begin{figure}
    \centering
    \includegraphics[width=0.85\columnwidth]{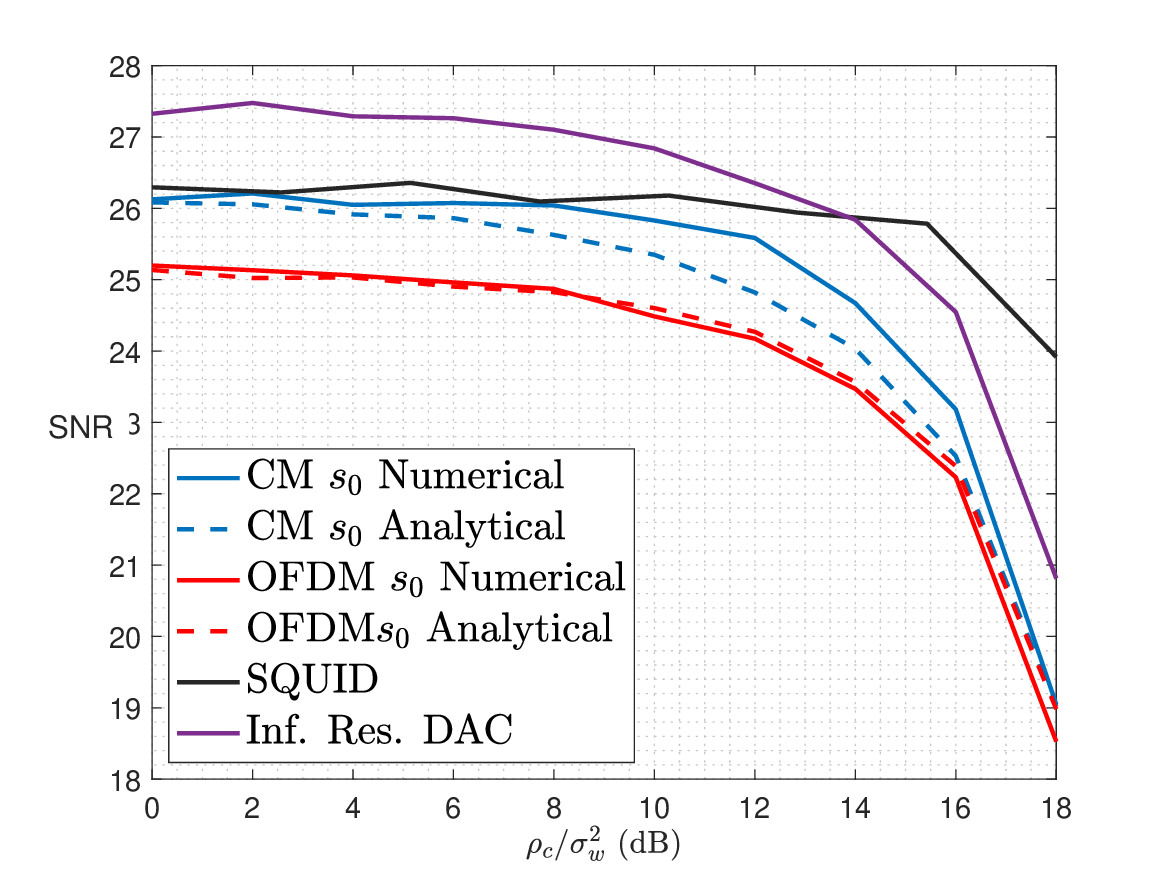}
    \caption{Sensing SNR (y-axis) vs. communication SNR (x-axis).}
    \label{fig:CSNRvsSNRK8}
\end{figure}
\section{SQUID-Based ISAC Waveform Design}

In the previous section, we explored the first approach, which involved designing the transmit waveform under the assumption of infinite-resolution DACs. In this section, we will delve into the second approach, specifically tailored for low-bit DACs. In this method, we construct the transmit waveform using a SQUID method to jointly optimize communication and sensing performance. We let the quantized signal ${\bf z}[n]$ be transmitted from the BS and designed using nonlinear quantized precoding techniques. We also want to have a desired gain towards the target location that is characterized by the transmit steering vector ${\bf a}_{M_{\rm t}}(\varphi_0)$. We denote the desired amplitude gain by ${\bf \alpha}$, i.e., $|{\bf a}_{M_{\rm t}}^{\Htran}(\varphi_0){\bf z}[n]|\geq\alpha$. The latter requirement can be expressed as ${\bf a}_{M_{\rm t}}^{\Htran}(\varphi_0){\bf z}[n]\gamma=\alpha$, where $\gamma\in \mathbb{C}$ and $|\gamma|\leq 1$. Note that in order to satisfy the sensing SNR constraint, the desired amplitude gain can be adjusted proportionally to the sensing power, $\alpha \propto \rho_{\rm s}$. Injecting this constraint into the penalty, we can write the optimization problem at time instance $n$ as
\begin{subequations}
\begin{align}
    \underset{\vect{z}[n],\beta,\gamma}{\text{minimize}} \quad  &\left\Vert {\bf s}[n]-\beta{\bf H}{\bf z}[n]\right\Vert^2+\sigma_w^2\beta^2\nonumber\\
    &+\left|{\bf a}_{M_{\rm t}}^{\Htran}(\varphi_0){\bf z}[n]\gamma-\alpha\right|^2 \\
    \text{subject to}\quad & z_m[n]=\pm \sqrt{\frac{P}{2M_t}}\pm \imagunit\sqrt{\frac{P}{2M_t}}, \quad m=1,\ldots,M_t, \label{eq:one-bit-DAC-constraint} \\
    &\beta>0,\quad |\gamma|\leq 1,
\end{align}
\end{subequations}
where $\beta$ denotes the common scaling factor adjusted at the BS as detailed in \cite{7967843}. After solving the optimization problem, the scaling factor corresponding to each UE may differ from the one optimized at the BS, as it cannot be estimated directly at the UEs for decoding purposes. For a constant-modulus constellation, the choice of the scaling factor does not impact the decoding of symbols by the UEs. 

We propose an iterative framework to solve the above problem. First, for a given $\gamma$, we solve the problem to find ${\bf z}[n]$ and $\beta$. The corresponding problem can be solved using the SQUID algorithm proposed in \cite{7967843}. 
 Next, we consider the optimization of $\gamma$, which is different than the considered problem in \cite{7967843} since we include the optimization of $\gamma$, which is related to beampattern design and does not exist in the works related to quantized precoding. 

   The optimization problem for a given ${\bf z}[n]$ with respect to $\gamma$ is given as
  \begin{subequations}
     \begin{align}
\underset{\gamma}{\text{minimize}} \quad & \left\vert {\bf a}_{M_{\rm t}}^{\Htran}(\varphi_0){\bf z}[n]\gamma-\alpha \right\vert^2 \\
\text{subject to } \,\,\, & |\gamma|\leq 1,
     \end{align} 
     whose solution is given as
     \begin{align}
         \gamma = \min\left( 1,\frac{\alpha}{\left| {\bf a}^{\Htran}_{M_{\rm t}}(\varphi_0){\bf z}[n] \right|}\right)e^{-\imagunit\arg\left({\bf a}_{M_{\rm t}}^{\Htran}(\varphi_0){\bf z}[n]\right)}.
     \end{align}
  \end{subequations}
The iterative updates of $\mathbf{z}[n]$, $\beta$, and $\gamma$ continue until convergence is achieved.
  
\section{Simulation Results}
In this section, numerical results are presented to show the effects of the quantization noise at the transmitter side. In the considered scenario, the BS is equipped with $M_{\rm t}= 128$ transmit and $M_{\rm r}=128$ receive antennas, and the transmit power is $100 \, {\rm W}$. The length of the transmitted block is selected as $L=64$. We consider a LoS channel between the BS and the communication UEs. To model the fading coefficients, we employ the urban macro-cell pathloss model as defined in \cite{3GPP_ch_model}. The bandwidth of the transmitted signal is $20$\,MHz, the noise variance is set to $-94$\,dBm, and the RCS variance is $\sigma_{\rm RCS}^2 = 1\,{\rm m}^2$. The two-way pathloss for the target signal is determined via the radar range equation as $\lvert \kappa_0 \lvert^2= \frac{c^2}{(4\pi)^3f_c^2 d_0^4}$, where $f_c=28$\,GHz is the carrier frequency, $d_0 = 400 \, {\rm m}$ is the distance from the target to the BS.

Firstly, the sensing SNR and BER of the ISAC system with 1-bit DACs are investigated. In this scenario, the number of users is fixed to $K=8$, and the distance between the users and the BS is $d_k = 500 \, {\rm m}$, and AoDs are generated randomly such that they do not overlap, and the distance between the target and the BS is $d_0 = 400 \, {\rm m}$. In addition, the analytically obtained SNR expression in \eqref{SNRFinal} is verified by the SNR obtained via Monte Carlo trials. Additionally, we also consider the sensing and communication performances of systems utilizing the OFDM-modulated signal for sensing signal $s_0$, and we evaluated the performance of the proposed SQUID-based waveform design method. In Fig. \ref{fig:CSNRvsSNRK8}, the sensing SNR performance is evaluated for different communication SNR $\rho_{\rm c}|\kappa|^2/\sigma_w^2$ values. To specify a common communication SNR, it is assumed that all users have the same channel gain. It can be observed that as the communication SNR increases, the sensing SNR decreases as expected. It can also be observed that the performance degradation due to 1-bit quantization is approximately $1 \, {\rm dB}$ compared to the ideal hardware case (infinite resolution DACs), where the OFDM waveform is utilized for radar signal $s_0$. The notable performance attained by low-resolution DACs can be attributed to the effective integration of linear signals through matched filtering operations. This process entails the coherent combination of intended signal power, resulting in a significant signal gain factor denoted as $L$. Also, it is noticeable that the analytical SNR value obtained by the proposed analysis framework closely approximates the system performance. In addition, it is observed that using the constant modulus (CM) sensing signal is beneficial in improving the sensing performance. When the CM modulation signal is employed, envelopes of the DAC inputs are also relatively constant; therefore, the quantization noise has lower significance. On the other hand, the OFDM waveform has significant envelope variations, which makes it more susceptible to nonlinear distortion. Consequently, using a dedicated sensing signal improves the sensing SNR performance by approximately $1 \, {\rm dB}$. It can also be seen that the waveform generation method via the SQUID method achieves slightly better sensing performance for lower SNR values. However, it requires higher computational complexity to generate the $M_{\rm t} \times L$ dimensional waveform matrix ${\bf Z} = [{\bf z}[0], \,\ldots,\, {\bf z}[L-1]]$. Hence, we can conclude that a heuristic waveform design using a CM sensing signal offers a practical solution characterized by a favorable tradeoff between performance and complexity considerations.

\begin{figure}
    \centering
    \includegraphics[width=0.85\columnwidth]{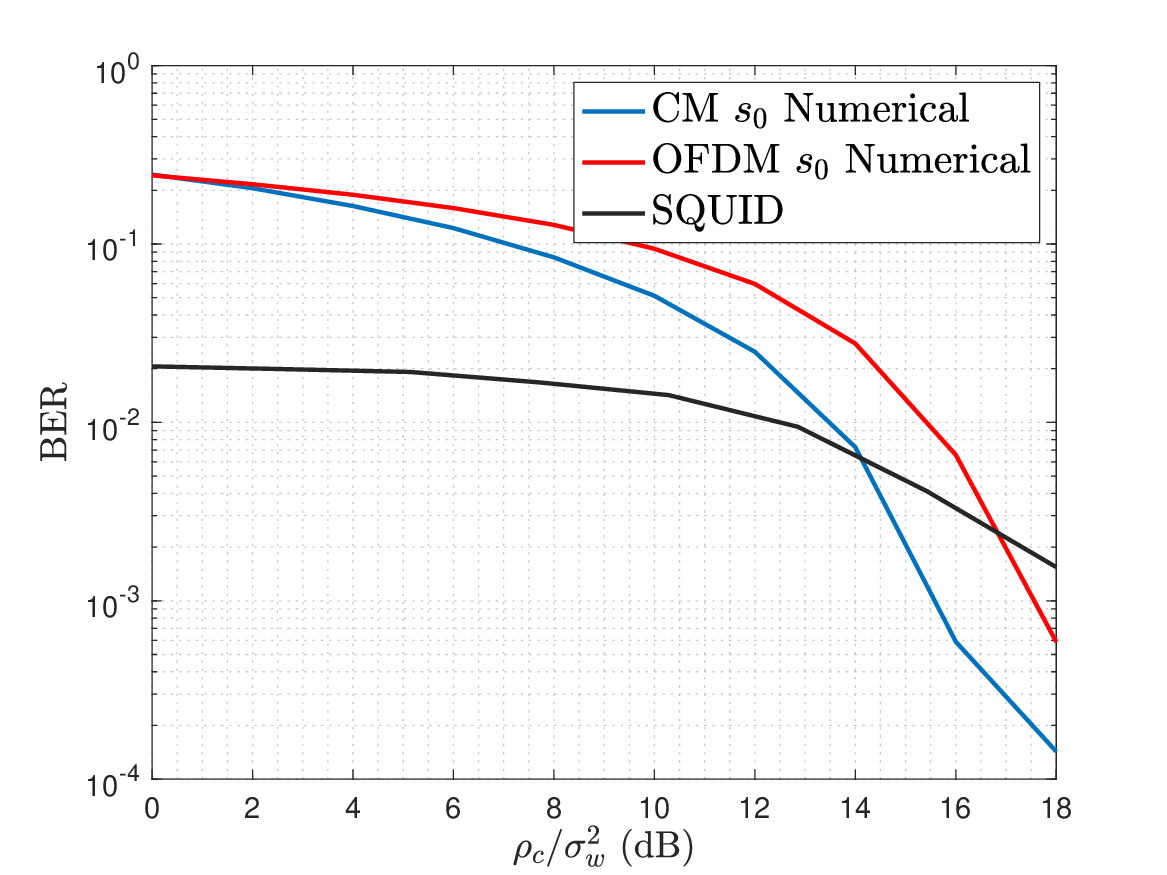}
    \caption{BER vs. communication SNR.}
    \label{fig:CSNRvsBERK8}
\end{figure}

In addition, uncoded BER performances are presented in Fig. \ref{fig:CSNRvsBERK8}, where QPSK modulated symbols with $L=64$ subcarriers are used. It can be seen that 1-bit quantization has a more significant effect on communication system performance than detection performance, and BER performance is significantly degraded compared to the ideal hardware. It can be observed that the use of the CM sensing signal improves the BER performance compared to the OFDM sensing waveform by approximately $2$\,dB. Furthermore, it is important to note that SQUID-based waveform generation significantly improves the BER performance compared to the quantization-unaware methods for lower SNR cases. Hence, we can conclude that the SQUID-based waveform generation provides superior joint sensing and communication performance for both tasks with increased computational complexity.

\begin{figure}
    \centering
    \includegraphics[width=0.85\columnwidth]{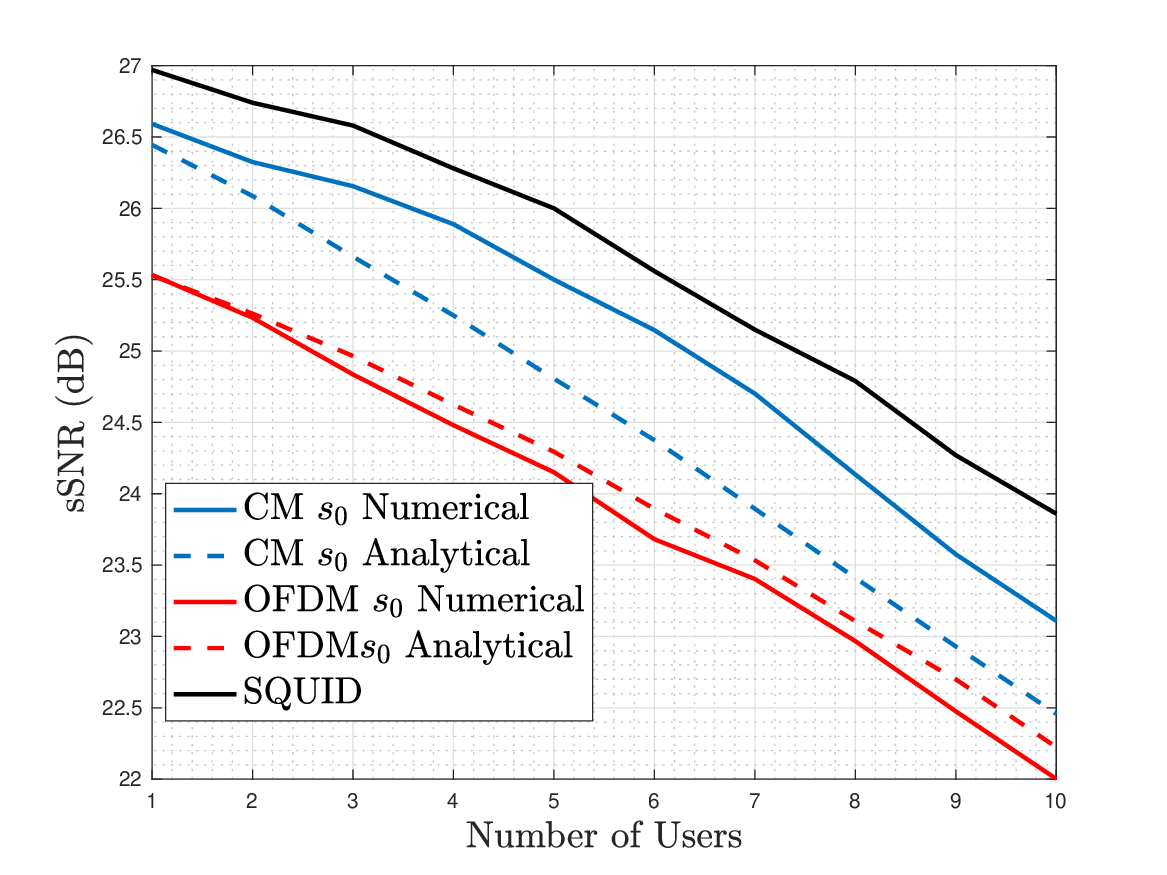}
    \caption{Sensing SNR (y-axis) vs. number of users (x-axis).}
    \label{fig:CSNRvsK}
\end{figure}

Fig. \ref{fig:CSNRvsK} delves into the impact of varying the number of users on the sensing SNR and uncoded BER performances. In these evaluations, the communication SNR per user remains fixed at $6 \, {\rm dB}$. As depicted in Fig. \ref{fig:CSNRvsK}, it becomes apparent that with an increasing number of users, the sensing SNR experiences a decline. This trend is ascribed to the proportional reduction in the allocated power specifically designated for sensing tasks, necessitated to uphold the communication SNR at its designated level. The analytical SNR curves for the OFDM sensing signal align more accurately with numerical results compared to the CM sensing signal. This is because the OFDM sensing signal produces a Gaussian distribution for the input of the quantization function, validating \eqref{eq:Czz}. In contrast, the CM sensing signal deviates from this Gaussian distribution, leading to discrepancies between the analytical and numerical curves. Nonetheless, the proposed analysis offers sufficient accuracy for system design.

\begin{figure}
    \centering
    \includegraphics[width=0.85\columnwidth]{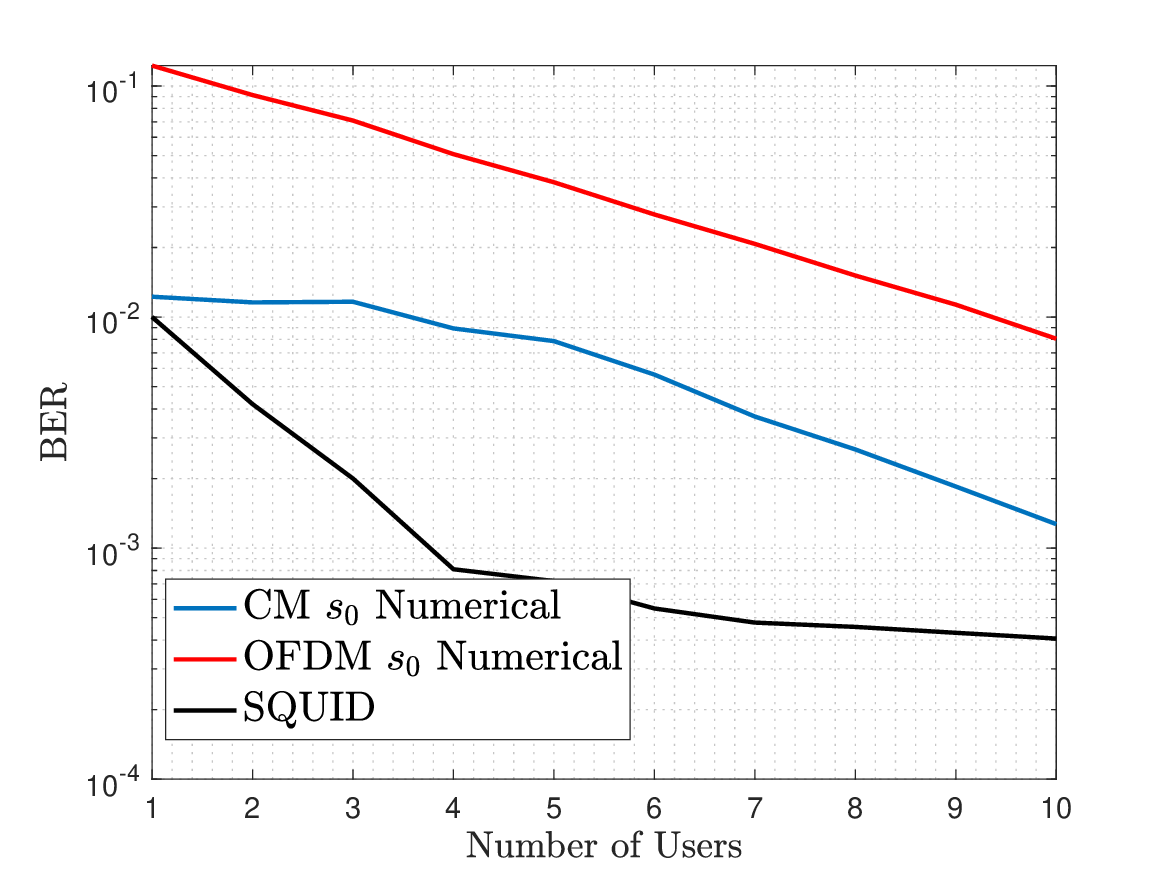}
    \caption{BER vs. number of users.}
    \label{fig:BERvsK}
\end{figure}

Conversely, despite this reduction in sensing SNR, discernible performance enhancements persist due to the adoption of CM sensing waveforms, in contrast to conventional OFDM waveforms. Fig. \ref{fig:CSNRvsK} also illustrates the superior sensing SNR metrics achieved by the SQUID technique. Notably, it is observed from Fig. \ref{fig:BERvsK} that SQUID attains comparable BER performance, evident in scenarios with a lower number of users, where heightened sensing SNR performance is prevalent.

Furthermore, it is noteworthy that the utilization of CM sensing signals exhibits improvements in both SNR and BER performance when juxtaposed with OFDM-modulated sensing signals. This observation underscores the efficacy of CM sensing waveforms in augmenting system performance across multiple evaluation metrics.

\section{Conclusion}

This study provides a meticulous analysis of downlink ISAC systems that are equipped with 1-bit DACs. We develop an analytical framework based on the Bussgang decomposition to assess the sensing SNR performance, providing valuable insights for system design. Introducing a heuristic waveform design methodology utilizing dedicated CM signals, we demonstrate through numerical analysis the effectiveness of improving sensing SNR performance. In addition, an optimization framework is proposed that jointly optimizes sensing and symbol detection performance, leveraging a beampattern constraint to enable efficient implementation of the SQUID algorithm. Through numerical evaluations, we illustrate tangible performance enhancements achieved with our proposed methodologies, contributing to the advancement of ISAC system design and optimization.

\bibliographystyle{IEEEtran}
\bibliography{IEEEabrv,refs}

\begin{thebibliography}{10}
\providecommand{\url}[1]{#1}
\csname url@samestyle\endcsname
\providecommand{\newblock}{\relax}
\providecommand{\bibinfo}[2]{#2}
\providecommand{\BIBentrySTDinterwordspacing}{\spaceskip=0pt\relax}
\providecommand{\BIBentryALTinterwordstretchfactor}{4}
\providecommand{\BIBentryALTinterwordspacing}{\spaceskip=\fontdimen2\font plus
\BIBentryALTinterwordstretchfactor\fontdimen3\font minus
  \fontdimen4\font\relax}
\providecommand{\BIBforeignlanguage}[2]{{%
\expandafter\ifx\csname l@#1\endcsname\relax
\typeout{** WARNING: IEEEtran.bst: No hyphenation pattern has been}%
\typeout{** loaded for the language `#1'. Using the pattern for}%
\typeout{** the default language instead.}%
\else
\language=\csname l@#1\endcsname
\fi
#2}}
\providecommand{\BIBdecl}{\relax}
\BIBdecl

\bibitem{10217169}
H.~Hua, T.~X. Han, and J.~Xu, ``{MIMO} integrated sensing and communication:
  {CRB}-rate tradeoff,'' \emph{IEEE Trans. Wirel. Commun.}, 2023.

\bibitem{10373185}
A.~Bazzi and M.~Chafii, ``Secure full duplex integrated sensing and
  communications,'' \emph{IEEE Trans. Inf. Forensics Security}, vol.~19, pp.
  2082--2097, 2024.

\bibitem{9737357}
F.~Liu, Y.~Cui, C.~Masouros, J.~Xu, T.~X. Han, Y.~C. Eldar, and S.~Buzzi,
  ``Integrated sensing and communications: Toward dual-functional wireless
  networks for {6G} and beyond,'' \emph{IEEE J. Sel. Areas Commun.}, vol.~40,
  no.~6, pp. 1728--1767, 2022.

\bibitem{10214237}
N.~T. Nguyen, N.~Shlezinger, Y.~C. Eldar, and M.~Juntti, ``Multiuser {MIMO}
  wideband joint communications and sensing system with subcarrier
  allocation,'' \emph{IEEE Trans. Signal Process.}, vol.~71, pp. 2997--3013,
  2023.

\bibitem{10124714}
C.~Wen, Y.~Huang, and T.~N. Davidson, ``Efficient transceiver design for {MIMO}
  dual-function radar-communication systems,'' \emph{IEEE Trans. Signal
  Process.}, vol.~71, pp. 1786--1801, 2023.

\bibitem{6891254}
E.~Björnson, J.~Hoydis, M.~Kountouris, and M.~Debbah, ``Massive {MIMO} systems
  with non-ideal hardware: Energy efficiency, estimation, and capacity
  limits,'' \emph{IEEE Trans. Inf. Theory}, vol.~60, no.~11, pp. 7112--7139,
  2014.

\bibitem{7967843}
S.~Jacobsson, G.~Durisi, M.~Coldrey, T.~Goldstein, and C.~Studer, ``Quantized
  precoding for massive {MU-MIMO},'' \emph{IEEE Trans. Commun.}, vol.~65,
  no.~11, pp. 4670--4684, 2017.

\bibitem{9780031}
M.~Deng, Z.~Cheng, L.~Wu, B.~Shankar, and Z.~He, ``One-bit {ADCs/DACs} based
  {MIMO} radar: Performance analysis and joint design,'' \emph{IEEE Trans.
  Signal Process.}, vol.~70, pp. 2609--2624, 2022.

\bibitem{10210086}
X.~Chen, L.~Huang, H.~Zhou, Q.~Li, and K.-B. Yu, ``Performance analysis of
  one-bit digital beamforming,'' \emph{IEEE Trans. Aerosp. Electron. Syst.},
  vol.~59, no.~6, pp. 8235--8245, 2023.

\bibitem{10064130}
Z.~Cheng, L.~Wu, B.~Wang, J.~Xie, and H.~Li, ``Relative entropy-based
  constant-envelope beamforming for target detection in large-scale {MIMO}
  radar with low-resolution {ADCs},'' \emph{IEEE Trans. Veh. Technol.},
  vol.~72, no.~8, pp. 10\,090--10\,106, 2023.

\bibitem{7894211}
S.~Jacobsson, G.~Durisi, M.~Coldrey, U.~Gustavsson, and C.~Studer, ``Throughput
  analysis of massive {MIMO} uplink with low-resolution {ADCs},'' \emph{IEEE
  Trans. Wirel. Commun.}, vol.~16, no.~6, pp. 4038--4051, 2017.

\bibitem{9096379}
X.~Yuan and J.~Zheng, ``Nonlinear one-bit precoding for massive {MIMO} downlink
  systems with $l_0$-norm constraint,'' \emph{IEEE Wireless Commun. Lett.},
  vol.~9, no.~9, pp. 1514--1518, 2020.

\bibitem{10043013}
Y.~Wang and A.~Li, ``{ADMM} based interference exploitation multi-user one-bit
  massive {MIMO} precoding,'' \emph{IEEE Trans. Veh. Technol.}, vol.~72, no.~7,
  pp. 9561--9566, 2023.

\bibitem{9399801}
Z.~Cheng, S.~Shi, Z.~He, and B.~Liao, ``Transmit sequence design for
  dual-function radar-communication system with one-bit {DACs},'' \emph{IEEE
  Trans. Wirel. Commun.}, vol.~20, no.~9, pp. 5846--5860, 2021.

\bibitem{9724233}
X.~Yu, Q.~Yang, Z.~Xiao, H.~Chen, V.~Havyarimana, and Z.~Han, ``A precoding
  approach for dual-functional radar-communication system with one-bit
  {DACs},'' \emph{IEEE J. Sel. Areas Commun.}, vol.~40, no.~6, pp. 1965--1977,
  2022.

\bibitem{9307295}
{\"{O}}.~T. Demir and E.~Bj\"ornson, ``The {Bussgang} decomposition of
  nonlinear systems: Basic theory and {MIMO} extensions [lecture notes],''
  \emph{IEEE Signal Process. Mag.}, vol.~38, no.~1, pp. 131--136, 2021.

\bibitem{3GPP_ch_model}
``{3GPP TR36.814 3rd Generation Partnership Project; Technical Specification
  Group Radio Access Network; Evolved Universal Terrestrial Radio Access
  (E-UTRA); Further advancements for E-UTRA physical layer aspects (Release 9)
  },'' 3GPP Std., Rev. V9.2.0, Tech. Rep., 2017.

\end{thebibliography}

\end{document}